\documentclass[%
reprint,
showpacs,preprintnumbers,
amsmath,amssymb,
aps,pre, 
nofootinbib,
longbibliography, 
floatfix,
]{revtex4-1}

\usepackage{enumitem}  
\usepackage{graphicx}% Include figure files
\usepackage{epstopdf}
\usepackage{dcolumn}% Align table columns on decimal point
\usepackage{bm}% bold math
\usepackage{hyperref}% add hypertext capabilities
\usepackage{xcolor,soul}
\usepackage{lipsum}  
\usepackage[none]{hyphenat}

\newcommand{\ie}{\textit{i.e.},\ }
\newcommand{\eg}{\textit{e.g.},\ }
\newcommand{\md}{\mathrm{d}}

\newcommand{\bi}{\mathrm{bi}}

\begin{document}

%\preprint{APS/123-QED}

\title{Creating localized plasma wave by ionization of doped semiconductors }
\author{Kenan Qu}
\author{Nathaniel J. Fisch}
\affiliation{%
	Department of Astrophysical Sciences, Princeton University,  Princeton, New Jersey 08544, USA
}%   

\date{\today}% It is always \today, today,
% but any date may be explicitly specified

\begin{abstract}
	Localized plasma waves can be generated by suddenly ionizing extrinsic semiconductors with spatially periodic dopant densities. The built-in electrostatic potentials at the metallurgical junctions, combined with electron density ripples, offer the exact initial condition for exciting long-lasting plasma waves upon ionization. This method can create plasma waves with a frequency between a few terahertz to sub-petahertz without substantial damping. 
	The lingering plasma waves can seed backward Raman amplification in a wide range of resonance frequencies up to the extreme ultraviolet regime. Chirped wave vectors and curved wave fronts allow focusing the amplified beam in both longitudinal and transverse dimensions. 
	The main limitation to this method appears to be obtaining sufficiently low plasma density from solid-state materials to avoid collisional damping.  
\end{abstract}

%\pacs{  }% PACS, the Physics and Astronomy
\maketitle

 \section{introduction}  \label{intro}
Localized plasma wave, because they linger in plasma, can potentially manipulate intense lasers in a variety of applications. Examples include replacing a seed laser pulse~\cite{Kenan_PRL2017} for initiating backward Raman amplification~\cite{Malkin_PRL1999} to avoid the technological challenges of frequency-shifting and synchronizing laser pulses, and heating hot electrons in compressing plasma~\cite{Schmit_PRL2010, Schmit_PRL2012} for the purpose of inertial confinement fusion. 
Localized plasma waves can also be formed to create a plasma holograph~\cite{Dodin2002prl}. 
However, creation of localized plasma waves is difficult with traditional methods which exclusively use propagating pulses (either laser pulses~\cite{LWave_SRS, geddes2004high} or electron pulses~\cite{LWave_eBeam}). The issues of preparing the beating lasers at the precise plasma frequency and delivering them into plasma have impeded creation of plasma waves with short wavelengths and sharp wave fronts, which are necessary for certain applications such as Raman amplification.

%The propagating pulses cause spatially spread plasma waves which lack a sharp wave front and may lead to unwanted precursors for backward Raman amplification~\cite{Tsidulko_PRL2002}. The achievable plasma wavelength is limited to the micrometer range due to the availability of high-power lasers. Preparing the beating lasers at the precise plasma frequency and delivering them into plasma become difficult at high plasma densities. The  also limits the minimum achievable plasma wavelength to the micrometer regime. 
%Injecting laser pulses into high-density plasma itself is difficult in applications like heating hot electrons in compressing plasma~\cite{Schmit_PRL2010, Schmit_PRL2012} for purpose of manipulating wave energy in an inertial confinement fusion target. 

The limitations associated with using propagating pulses might be avoided by ionizing solid-state materials for generating plasma waves. Solid-state materials show flexible and precise controllability over the spatial density profile of the generated plasmas. These unique advantages combined with ionization enable dielectric laser acceleration of electrons~\cite{Acceleration_PoP2017, IonAccel_PRE2004, IonAccel_PRL2016}, protons, and heavy ions~\cite{IonAccel_PoP2018, IonAccel_PRE2004, IonAccel_PRL2016}. Solid-state density plasma also offers an electrostatic potential in crystal channels to confine positively charged particles and accelerate them to energy near $300\,\mathrm{GeV}$~\cite{DodinPoP2008, ShinPRAB2016, SHIN201594, ShinAPL2014}. 
Ionizing micro-scale nanowire arrays using strong lasers~\cite{Rocca2015, Rocca2016, Rocca2018} can create ultra-dense plasmas for bright ultrafast quasi-monoenergetic neutron point sources and nuclear fusion. Solid-state effects have also been adapted in designing plasma photonic crystals~\cite{Lehmann_PRL2016, Lehmann_PoP2017} for frequency-selective reflection of high-intensity lasers, and in x-ray Raman amplifiers for suppressing parasitic and modulational instabilities~\cite{MalkinPRL2007}. However, in generating plasma waves, directly ionizing solid-state materials with periodic densities leads to a nonhomogeneous ion density, which eventually causes strong plasma wave damping, as we will show. Creating lingering plasma waves would be best accomplished with homogeneous ion density but periodically varied electron densities. 

In this paper, we identify a new possibility for creating plasma waves with precision, whereas satisfying the homogeneity requirements, namely, by using periodically doped semiconductors~\cite{hu2010modern}, which have electron density ripples and built-in electrostatic fields in a homogeneous ion background. When ionized, they offer the exact initial condition for initiating high-amplitude plasma waves. The energy to initiate plasma waves is provided by the alternating high and low chemical potentials arising from periodic doping. Sudden ionization releases the chemical potential energies to drive the oscillations of the liberated electrons, thereby generating strong localized plasma waves. Since the ion density is homogeneous, the generated plasma waves do not suffer other kinetic effects due to inhomogeneity. Collisional damping in the cold plasma can be mitigated when the plasma frequency is on the terahertz to subpetahertz range. 

\begin{figure}[b]
	\includegraphics[width=0.9\linewidth]{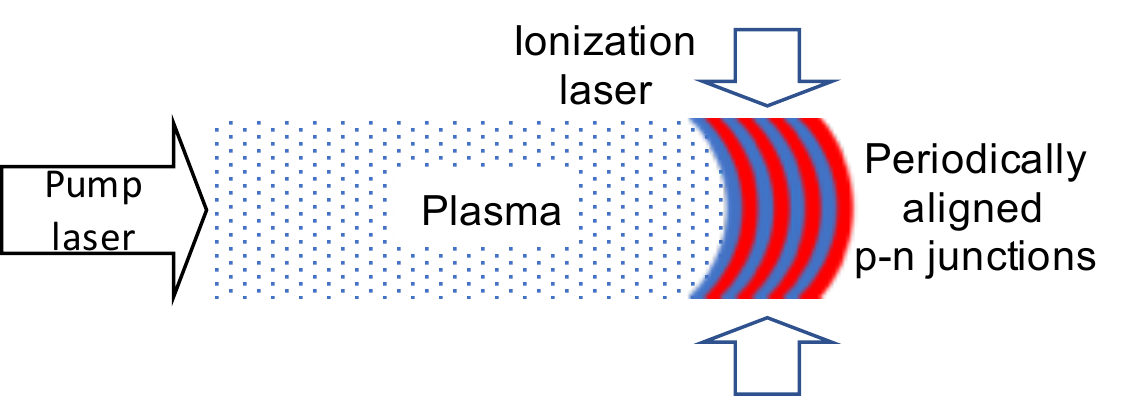}
	\caption{Schematics for generation of a plasma waveby suddenly ionizing periodically aligned p-n junctions and its implementation in backward Raman amplification. The curved shape of the p-n junctions demonstrates manipulation of the plasma wave front can create a pre-focused backscattered pulse.   }
	\label{fig0}
\end{figure}

An important application would be if the generated plasma wave can replace the laser seed in backward Raman amplification. This might be envisioned as follows: To create the plasma wave seed, a periodically doped semiconductor with the appropriate wave number is placed at the end of a plasma channel, as illustrated in Fig.~\ref{fig0}.  An external laser ionizes the semiconductor to the plasma density which matches the density of the plasma channel. The ionization also initiates the plasma wave which persists. When the Raman pump laser propagates through the plasma channel and reaches the plasma wave, it produces a frequency-downshifted reflection beam. The reflection beam can have the sharpness for efficient Raman compression~\cite{Kenan_PRL2017, Tsidulko_PRL2002}. The wave front and the wave numbers of the plasma wave can also be curved and chirped by design, which allows focusing the amplified beam in both longitudinal and transverse directions.

The paper is organized as follows: In Sec.~\ref{creation}, we first describe the built-in electrostatic field in a periodically doped semiconductor. We then show that sudden ionization of the semiconductor initiates a strong plasma wave motion. We subsequently discuss the effects of inhomogeneous plasma density and collisional damping. In Sec.~\ref{impl}, we analyze the possibility and advantages of implementing the created plasma wave in initiating Raman amplifications. We find the parameter regimes of the plasma density and wave number that are suitable for Raman amplification. In Sec.~\ref{concl}, we conclude the paper and address potential challenges associated with this technique. We also offer more possible applications that can take advantage of the localized plasma waves.

\section{Creating plasma waves} \label{creation}

To quantitatively explain how a plasma wave is generated by ionization, we first analyze the periodic chemical potential in doped semiconductors. The Fermi level, which describes the occupation of the charge carriers, lies in the middle of the band-gap of intrinsic semiconductors. It shifts if the semiconductor is doped with impurity atoms due to electronic band mismatch.  Specifically, doping impurities that are rich in/deficient of valence electrons shifts the Fermi level towards the top/bottom of the band-gap, forming an n-type/p-type semiconductor, and the change in Fermi levels are 
\begin{align}
\delta V_\mathrm{Fn} &= (K_BT/e) \ln(n_n/N_C), \label{01} \\
\delta V_\mathrm{Fp} &= (K_BT/e) \ln(N_V/n_p), \label{02}
\end{align}
respectively, where $K_B$ is the Boltzmann constant, $T$ is the temperature, $e$ is the natural charge, $N_{C,V}$ is the effective densities of states of the conduction/valence band, and $n_{n,p}$ are the dopant density of electrons/holes. 
%Since intrinsic semiconductors have low concentrations near $N_{C,V}=10^{18}{\sim} 10^{19}\, \mathrm{cm}^{-3}$, large Fermi level shifts can be achieved with low doping densities. 

To create the initial condition for plasma waves, n-type and p-type semiconductors are fabricated adjacently, as illustrated in Fig.~\ref{fig0}. The spatially varying Fermi level induces a periodic internal electrostatic potential $V_\bi {=} \delta V_\mathrm{Fn} {+} \delta V_\mathrm{Fp} $ at the metallurgical p-n junctions. The potential is associated with a periodic electrostatic field whose amplitude is 
\begin{equation}\label{03} 
E = \nabla V_\bi. 
\end{equation} 
The field causes the charged carriers to diffuse across the p-n junctions, thereby forming an electron density ripple $\delta n_e {=} (\varepsilon_0/e) \nabla E$, where $\varepsilon_0$ is the permittivity of vacuum.

Note that periodically doped semiconductors are fundamentally different from other solid-state materials for their ability to support large built-in electrostatic fields while maintaining homogeneous ion density. This is because the semiconductor chemical potential is induced by a low density of dopant atoms, rather than relying on a large gradient pressure of nonhomogeneous plasma density. Since the intrinsic semiconductors have much lower density of states ($N_{C,V}=10^{18}{\sim} 10^{19}\, \mathrm{cm}^{-3}$) than the atom electron density (${\sim} 10^{23}\, \mathrm{cm}^{-3}$),  a dopant density near $N_{C,V}$ can induce a large shift of Fermi level without significantly changing the total atomic density.

Plasma waves can be excited in the periodically doped semiconductor by sudden ionization using a strong laser. The laser intensity controls the degree of ionization and hence the plasma density~\cite{Suckewer_2002_upshift, Suckewer_2005_upshift, Nishida2012}. Since photoionization takes place in a laser cycle, the atomic density profile and the built-in electrostatic field of the semiconductor is maintained. The liberated electrons then oscillate, driven by the electrostatic field. A simple hydrodynamic model describes the evolution of plasma dynamics 
\begin{align}
	\partial_t n_e &= - \nabla \cdot(n_e {u}_e), \label{11} \\
	m_e\partial_t (n_e{u}_e)  &= -e n_e {E}  - 3 K_BT \nabla n_e, \label{12} \\
	\nabla\cdot{E} &= (e/\varepsilon_0)(n_e-Zn_i), \label{13}
\end{align}
where $n_{e,i}$ is the electron/ion density, $u_e$ is the electron flow velocity, $m_e$ is the mass of an electron, and $Ze$ is the ion charge. In a general form, the electron density and electric field can be written as $n_e = \bar{n}_e + \delta n_e$ and $E = E_0+\delta E$ where $\bar{n}_e ({=}Zn_i)$ and $E_0$ denote the steady-state values, and $\delta n_e$ and $\delta E$ denote the perturbation. Then, we obtain a wave equation
\begin{multline} \label{14}
	\partial_{tt} \delta n_e - \frac{e}{m_e}[\nabla(E_0\delta n_e)  + (\nabla\bar{n}_e)\delta E] \\
	- \frac{e^2\bar{n}_e}{\varepsilon_0m_e}\delta n - \frac{3K_BT}{m_e} \nabla^2 \delta n_e =0,
\end{multline}
and $\nabla\delta E = (e/\varepsilon_0)\delta n_e$. If the ion density is homogeneous, \ie $\nabla\bar{n}_e = E_0 =0$, then Eq.~(\ref{14}) reduces to the standard wave equation for plasma oscillation with the plasma frequency $\omega_p = \sqrt{e^2\bar{n}_e/(\varepsilon_0m_e)}$. If the ion density is nonhomogeneous, the plasma frequency becomes position dependent and the first-order spatial derivative terms in Eq.~(\ref{14}) induce damping of the wave energy.

Exciting coherent electron oscillations can be achieved, from Eq.~(\ref{12}), by preparing either an initial periodic electrostatic field or an initial density gradient at a finite temperature. A periodically doped semiconductor with homogeneous ion density creates a superposition of two eigenmodes of plasma waves with equal wave vector, frequency, and amplitude, but the opposite phase velocities. The energy density of each plasma wave is identically $\varepsilon_0 \delta E^2/2$. Since they are eigenmodes of the homogeneous plasma, this lingering localized plasma wave only decays at the Landau damping rate.

%Importantly, the periodic spatial distribution of $E$ and $\delta n_e$ in a homogeneous plasma background offer a stable initial condition for exciting the plasma wave when the extrinsic semiconductor is suddenly ionized. Upon ionization, the liberated electrons are accelerated by the internal electric field. Since all the electrons start to move from rest, their motions are synchronized and form a coherent plasma oscillation. 
%An initial density gradient for exciting plasma waves can be provided by ionization of materials with alternating ion densities. Since it does not have built-in electrostatic fields, 

To appreciate the advantage of creating plasma waves by ionization of semiconductors, we compare it to ionization of solid-state materials without built-in electrostatic fields. In the latter case,  achieving the same plasma wave amplitude requires a large electron density gradient
\begin{equation}\label{21}
\delta E = -\frac{3K_BT}{e} \frac{\nabla n_e}{n_e}.
\end{equation}
The density nonhomogeneity increases exponentially with large wave amplitude. For example, creating a monochromatic plasma wave $\delta E{=}\bar{E}\sin(kz)$ in the $z$-direction with wave number $k$ requires an ion density 
\begin{equation}\label{22}
Zn_{i}(z) = n_{e0} \exp\left[ \frac{-e\bar{E}}{3K_BT}\cos(kz) \right],
\end{equation}
where  $n_{e0} $ is the mean electron density.

\begin{figure*}[thp]
	\includegraphics[width=0.9\linewidth]{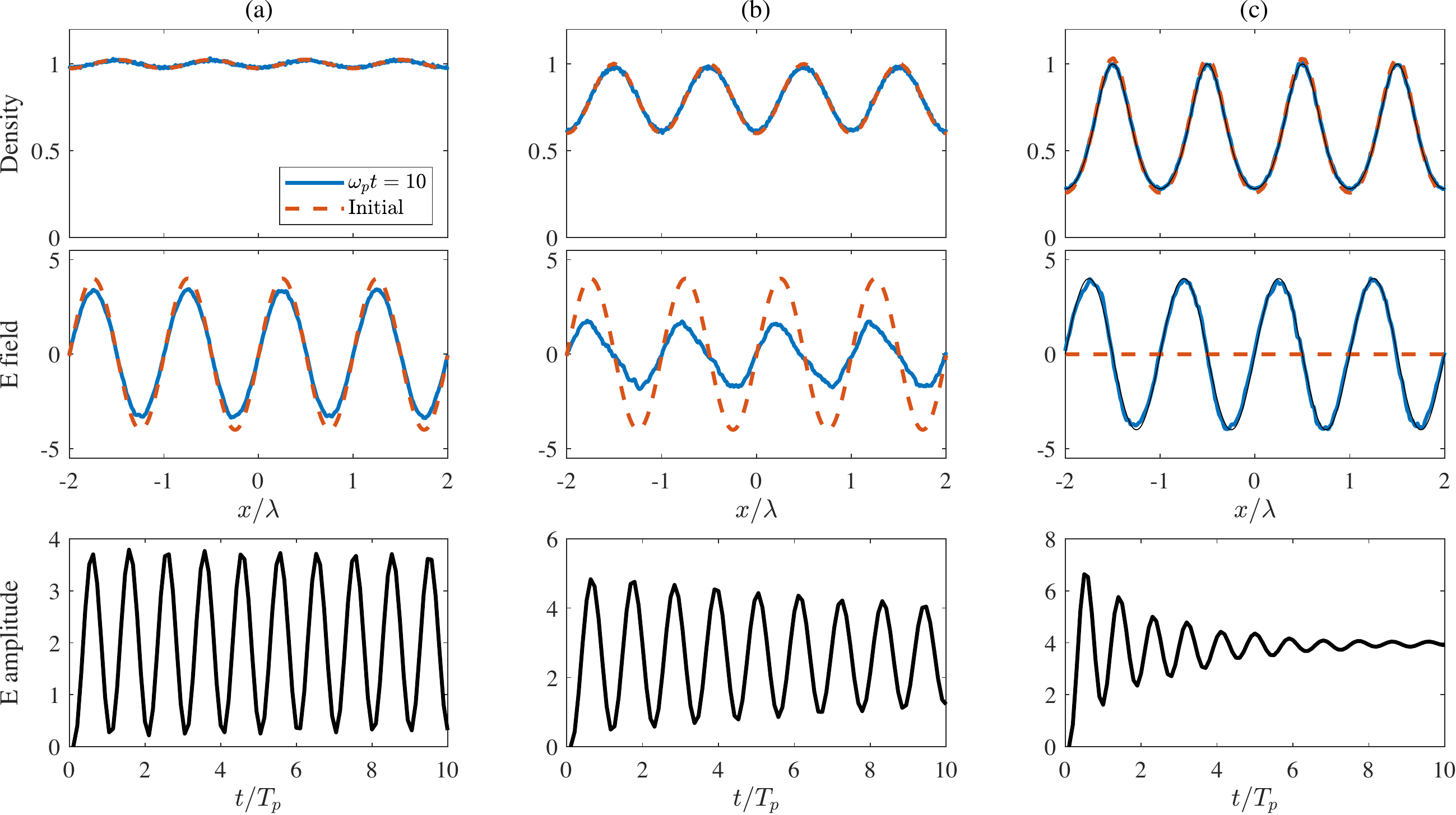}
	\caption{ Generation and evolution of plasma waves by ionization of periodically doped semiconductors with (a) homogeneous ion density and an initial electrostatic field, (b) ion density of $20\%$ nonhomogeneity and an initial electrostatic field,  and (c) materials having alternating electron density and no initial electrostatic fields. The density axis are normalized to $n_e {=} 1.12{\times} 10^{25}\, \mathrm{m}^{-3}$, and the fields are normalized to $1{\times} 10^{8}\, \mathrm{V/m}$. The black curves in (c) show the amplitude evolution of the sinusoidal field. Here, $\lambda{=}0.5\,\mu\mathrm{m}$ is the plasma wave wavelength and  $T_p$ is the period of plasma oscillation.}
	\label{fig1}
\end{figure*}

However, the nonhomogeneous plasma density in its steady state deleteriously causes strong wave damping, thereby limiting its ability. The large damping can be anticipated as follows. The density gradient causes electron diffusion and a position-dependent steady-state electrostatic potential $\Phi(z)$.  When the electrons are ionized, they oscillate coherently only immediately after ionization. Then they are accelerated to different kinetic energies depending on $\Phi(z)$. All electrons with lower kinetic energy than $-e\Phi$ get trapped inside the potential at different times: The slow electrons, which are originally in the high density region, get trapped first, whereas the electrons which move into the high density region are decelerated and then get trapped. These electrons are trapped in different trajectories with different bounce frequencies $\omega_B(z) {=} \sqrt{keE_0(z)/m}$. Since the electrostatic potential depth varies, the bounce periods of the trapped electrons are widely spread. Hence, they accumulate different phases within the time frame of $2\pi/\Delta\omega_B$ to smear out the coherent plasma oscillation.

Another damping effect in high-density plasma is due to collisions of electrons with ionized ions and with the unionized semiconductor lattice. Solid-state semiconductors have high atom densities. When completely ionized, the plasma becomes strongly coupled and plasma waves do not exist. The high collision rate between the ionized electrons and ions can be reduced by either increasing the electron temperature or reducing the plasma density with partial ionization. To estimate the maximum plasma density, we compare the electron-ion collision frequency which scales linearly with plasma density, \ie $\nu_{ei}{\sim} 10^{-11} (Z\bar{n}_e/\mathrm{m}^{-3}) (K_BT/ \mathrm{eV})^{-3/2}\,\mathrm{Hz}$ and the plasma frequency which  scales with $\omega_p {\sim}\bar{n}_e^{1/2}$. For silicon ($Z {=}14$), the maximum plasma density is $\bar{n}_e {\sim} 3{\times} 10^{19}\,\mathrm{cm}^{-3}$ at $20\,\mathrm{eV}$, and it increases to $3{\times} 10^{22}\,\mathrm{cm}^{-3}$ at $200\, \mathrm{eV}$. The corresponding plasma frequency is $2\pi {\times} 50\, \mathrm{THz}$ for $20\,\mathrm{eV}$ and $2\pi {\times} 0.5\, \mathrm{PHz}$ for $200\,\mathrm{eV}$, respectively.

At the instant of ionization, the ion temperature is three to four orders of magnitudes smaller than the electron temperature depending on the ion mass. The ion temperature remains low if the electron-ion collision is negligible. Since the lattice binding energy is tens of eV, ion-lattice collision will not damage the lattice and only slightly increases the lattice temperature. The concern for plasma wave damping is the collision between electrons with the lattice at finite temperatures. Fortunately, the electron-lattice collision is very rare due to the quantum nature of electrons. In a weakly ionized semiconductor, the electron relaxation due to the semiconductor lattice and impurity, known as the electron mobility, is typically above a few picoseconds for elemental and binary semiconductors~\cite{hu2010modern}, and is submicroseconds for organic semiconductors~\cite{Kahn3}. It is hence negligible when the plasma frequency is above a few terahertz. Therefore, the electron-ion collision rate and the electron-lattice collision rate set the limit of the achievable plasma frequency to a range between a few terahertz to sub-petahertz for plasma at the temperature of tens of or few hundreds of eV.

To show these effects, we numerically simulate the creation and evolution of the plasma waves generated by different methods using the full kinetic collisional code EPOCH~\cite{Arber2015contemporary}, and show the results in Fig.~\ref{fig1}. Plasma waves with the same amplitude $\bar{E}{=} 4{\times} 10^{8}\, \mathrm{V/m}$, wave number $k{=}2\pi/0.5\,\mu\mathrm{m}^{-1}$, and electron temperature $K_BT {=} 50\, \mathrm{eV}$ are generated from different initial conditions. In a homogeneous plasma [case (a)], the plasma wave, resulted from partially ionizing a periodically doped semiconductor, maintains its amplitude within the simulation time of ten plasma wave periods. For comparison, when a $40\%$ ion density modulation is added to the initial condition in case (b), the wave amplitude decays nearly $40\%$ within the same simulation time. The curve of its electric field reveals that a part of the wave energy is transfered to a high-$k$ mode after ten plasma wave periods.  

Without an initial electrostatic field, exciting plasma waves by nonhomogeneous plasma requires a steep density gradient, as shown in case (c). To create the same electric field, we use an ion density profile: $Zn_i {=} 0.59{\times} \exp[-0.637\cos(kz)]{\times}10^{25}\, \mathrm{m}^{-3}$. At time $\omega_p t{=}10$, although the electric field exhibits a spatial sinusoidal pattern, the field or the electron density no longer oscillate in time. The stationary spatial sinusoidal pattern is the electrostatic potential $\Phi_B$ caused by diffusion. The temporal amplitudes of the electrostatic field with wave vector $k$ shows that the plasma wave in nonhomogeneous plasma decays quickly at the rate of nearly $0.25\omega_p$. The decay rate is close to the bouncing frequency $\omega_B$, matching our model of kinetic wave decay. In contrast, the plasma wave in homogeneous plasma [case (a)] barely decays in the same time frame.

\section{Implementing in Raman amplification} \label{impl}

To implement the plasma wave seed for Raman amplification, one can place the periodically doped semiconductors in one end of a plasma channel, as illustrated in Fig.~\ref{fig0}. Ionization of the semiconductor is achieved using a separate laser with an intensity of $10^7 - 10^{10}\, \mathrm{W\,cm}^{-2}$ to ensure that the ionized plasma density matches the plasma channel~\cite{Suckewer_2002_upshift, Suckewer_2005_upshift, Nishida2012}. Ionization creates two plasma waves with oppositely directed phase velocities. However, they have essentially zero group velocity and hence will persist long enough for a high-power pump laser to excite backward Raman amplification. It was pointed out~\cite{Kenan_PRL2017} that using a plasma wave seed can avoid the experimental challenges of aligning and synchronizing two short laser pulses. With a proper wave vector, the plasma wave can reflect the pump beam and also downshift its frequency by $\omega_p$. The reflection beam compresses the pump laser pulse in the plasma channel through Raman instability and eventually yields high intensity output.

The advantage of using semiconductors for creating plasma waves  is that the wave vector $k$ is flexibly and precisely controlled with the junction separation, which enables amplification over a range of laser resonance frequencies. For triggering efficient Raman amplification, the requirement for the plasma wave seed is the phase matching condition $k = k_a {+} \sqrt{\omega_a(\omega_a {-} 2\omega_p)}/c$ for a certain laser with frequency $\omega_a$ and wave vector $k_a {=}\omega_a/c$ ($c$ is the speed of light).  With current semiconductor fabrication technology, the maximum wave vector $k_\mathrm{max}$ can reach  $2\pi/10\,\mathrm{nm}^{-1}$ in a silicon semiconductor, which is sufficiently large to amplify extreme ultraviolet (EUV) light. It  can further increase by using semiconductors with larger charge carrier mobility. 
% Importantly, the plasma wave can even convert a partially coherent light source into a coherent short pulse through the beam cleaning process~\cite{Kenan_incoherent2017}. 

A large plasma wave seed wave vector can also increase the seed amplitude, which is important for quickly reaching the high-efficiency pump-depletion regime. This can be shown from the asymptotic equivalence condition between a laser seed and the backscattering from a plasma wave seed~\cite{Kenan_PRL2017},
\begin{equation}\label{04} 
E_b(t,z) = \frac{\omega_b}{\omega_a} \frac{e}{2m_ec^2} \int \md z' E_a(t,z-z') E(z'), 
\end{equation} 
where $E_{a,b}$ is the electric field of the pump/backscattered beam, and $\omega_b$ is the frequency of the seed pulse or the backscattered beam. Here, we introduce the factor of $1/2$ because only one of the counterpropagating plasma waves can efficiently backscatter the pump. 
In the optical limit $\omega_a{\cong} \omega_b$, the plasma wave amplitude solely determines the scattering cross section $\sigma {=} eE/(2m_ec^2)$. According to Eq.~(\ref{03}), the value of $E$ is limited by $kV_\mathrm{bi}$. 
The internal electric potential $V_\bi$ is limited by the band-gap, which is between $0.18\,\mathrm{eV}$(InSb) and $3.4\,\mathrm{eV}$(GaN). 
For a silicon semiconductor with $V_\bi {\approx} 1\,\mathrm{V/m}$, the electric field is $E{\approx}1.3{\times}10^{7} \,\mathrm{V/m}$ when the doping period is $2\pi/k {=}0.5\,\mu \mathrm{m}$. Hence, the corresponding cross section $\sigma {=}(78\, \mathrm{mm})^{-1}$ indicates that it requires a $7.8$-mm-long plasma wave for the backscattered beam to reach only one-tenth the pump amplitude, which would be inefficient for seeding Raman amplification. However, the cross section $\sigma$ increases proportionally with a larger plasma wave vector. 
At $k_\mathrm{max} {\approx} 2\pi/10\,\mathrm{nm}^{-1}$, the electric field amplitude reaches $E{\approx} 5{\times} 10^{8}\, \mathrm{V/m}$, and the cross section reaches $\sigma{=}(2\, \mathrm{mm})^{-1}$. 
Further increasing $\sigma$ would require using semiconductors with larger band-gaps, \eg GaN or even diamond. 
However, we must emphasize that the plasma frequency and wave vector cannot arbitrary increase because their optimal values for a certain pump laser frequency are constrained by factors including the Raman growth rate, forward scattering, and filamentation in backward Raman amplification~\cite{Malkin2014}.

Moreover, apart from providing a strong plasma wave amplitude, ionization of an appropriately doped semiconductor can, in principle, exploit group velocity dispersion in the resulting plasma. Focusing in the pulse propagation direction can be achieved through self-contraction of a pre-chirped pulse in a high density plasma with strong group velocity dispersion ~\cite{PRL2012_Toroker, Kenan_PRL2017}. The proposal is to prepare a plasma wave which has an increasing $k$ in the laser propagation direction. When it interacts with a pump pulse, the low-$k$ components are scattered before the high-$k$ components, creating an up-chirped counter-propagating beam. The spectrum of the chirped plasma wave increases the spectral width of the backscattered pulse, thereby supporting a shorter pulse duration. The pulse contraction during its propagation in plasma increases its peak intensity and wave front sharpness.

For a pump laser pulse with duration $\tau_a$ and peak amplitude $E_a$, when scattered by a non-chirped plasma wave with length $L$ the backscattered pulse has a duration $\tau_b{=}\tau_a{+}L/c$ and peak amplitude is $E_b{=}E_a\sigma L$. If the plasma wave has a chirped spectrum with spectral width $\Delta\omega_p$, the spectral width is added to the backscattered pulse. After compression, the ultimate pulse duration becomes $\tau_b' {=} (1/\tau_b {+} \Delta\omega_p/\mathrm{TBP})^{-1}$, where TBP(${\sim1}$) refers to the time-bandwidth product. The peak amplitude is then increased to $(\tau_b/\tau_b')E_b = (1{+}\Delta\omega_p\tau_b/\mathrm{TBP}) E_a\sigma L$. For example, if $\Delta\omega_p\tau_b/\mathrm{TBP} {=} 2$, the pulse amplitude can be increased by a factor of~$3$. Note that, ideally, a chirped plasma wave requires the plasma frequency to be changed accordingly by means of introducing a plasma density gradient or varied magnetic fields. Nevertheless, the effect of frequency mismatch is negligible if the spectrum of the backscattered beam is narrow.

Focusing the backscattered beam in the transverse direction can be achieved through manipulating the plasma wave front. It takes advantage of the flexible geometrics of solid-state materials that the initial metallurgical junctions can be aligned in curved equiphase surfaces, as illustrated in Fig.~\ref{fig0}. When interacting with a plane-wave pump, the curved phase front produces a prefocused backscattered pulse as if a concave mirror. The focal point can be controlled with the wave fronts of the plasma wave and pump pulse together. 

Working with solid state, precise spatial localization and alignment of the electron density can be implemented, enabling flexible control of the generated plasma wave front and its sharpness~\cite{Leblanc2017, NatC_Vieira, PRL_Mendonca, Plasma_qplate}. The engineered plasma wave front allows creating three-dimensional structured light with orbital angular momentum~\cite{allen1992OAM}, nondiffracting beam size~\cite{BesselBeam}, and vector beams~\cite{VectorBeam}. 
More interestingly, the plasma wave front can be advantageously used to correct the laser phase distortion ~\cite{PC_compensation} when propagating through a nonhomogeneous plasma. Provided that the density and geometry of the plasma channel are known, the plasma wave seed wave front can be designed to perfectly match the wave front of the output pump beam. Then the backscattered beam becomes a near-phase conjugation (except for a frequency mismatch) of the pump, which reverses the phase distortion when propagating against the pump.

For experimental realizations, one can consider ionizing multi-junction semiconductors with alternative p- and n-layers. The materials are readily available as components of high-efficiency solar cells~\cite{YAMAGUCHI200578}. Creation of the curved junctions for purposes of focusing laser and compensating distortion can be achieved with organic semiconductors~\cite{Kahn1, Kahn2, Kahn3, Kahn4}, which are mechanically deformable. Organic semiconductors are made of carbon and hydrogen, and hence are also advantageous for producing plasma with light ions and less mobile electrons.

Although ionizing semiconductors can create large wave vectors to facilitate Raman amplification of short-wavelength pulses, the plasma frequency cannot exceed a few tens or hundreds of terahertz constrained by the high collisional damping at cold temperatures. The finite plasma frequency limits the Raman growth rate $\gamma \propto \sqrt{\omega_a\omega_p}$  at high pump frequencies. Assuming $\omega_p$ is one or two orders of magnitudes smaller than $\omega_a$ for achieving competitive growth rate, the pump is limited to the EUV frequencies. For example, the pump frequency is limited to $2\pi {\times} 5\, \mathrm{PHz}$ at $20\,\mathrm{eV}$, or $2\pi {\times} 50\, \mathrm{PHz}$ at $200\,\mathrm{eV}$.

\section{Conclusion and discussion} \label{concl}

In conclusion, we investigate the possibility of creating localized plasma waves by ionization of periodically doped semiconductors. The high-amplitude built-in electrostatic fields and electron density ripple can initiate strong plasma waves in a homogeneous density plasma. This method is advantageous in multiple aspects including: (i) high wave amplitude, (ii) long life time, (iii) large wave vector, and (iv) flexible structure of the wave fronts. Using current technology with silicon semiconductors, the generated plasma wave can reach an amplitude up to $\bar{E}{=} 4{\times}10^8\, \mathrm{V/m}$, life time of multiple picoseconds, and wave numbers of up to $2\pi/10\,\mathrm{nm}^{-1}$. These parameters can further increase when using different materials such as GaN or organic semiconductors. 
The dominant limiting factor is the electron-ion collisional damping rate, which determines the maximum achievable plasma density for Langmuir waves to linger at least a few wave periods. At an electron temperature of $20\,\mathrm{eV}$, the plasma density cannot exceed $3{\times} 10^{19}\,\mathrm{cm}^{-3}$ ($\omega_p {\sim2}\pi{\times} 50\, \mathrm{THz}$) to avoid strong collisional damping. The number can increase to $3{\times} 10^{22}\,\mathrm{cm}^{-3}$ ($\omega_p {\sim2}\pi{\times} 0.5\, \mathrm{PHz}$) at $200\,\mathrm{eV}$. 
This is sufficient for preparing a plasma wave seed for Raman amplification of $1\,\mu\mathrm{m}$ laser or EUV lights.

The limitation of the maximum plasma density poses the primary challenge in ionizing solid-state semiconductors because they typically have very high densities. For silicon, even only ionizing electrons in the outermost shell would create plasma frequencies on the order of $2\pi{\times}4 \,\mathrm{PHz}$. Organic semiconductors have lower atomic densities but they still generate plasma frequencies in the sub-petahertz range. 
In our discussion, we focus on using a moderately strong laser to control the plasma density by partial ionization. This setup requires the Raman pump laser to have a sufficiently long weak wave front which produces reflection without causing ionization. But too weak a wave front cannot produce a sufficiently intense reflection which is needed for efficient Raman compression. The trade-off between suppressing ionization and obtaining strong reflection might be mitigated if one can use the front of the Raman pump, instead of a separate ionization laser, to ionize the semiconductor. After creation, the plasma wave immediately interacts with the pump to backscatter pump energy. A pump front with precisely controlled intensity and duration would then ionize the plasma to the appropriate density. This method may allow the Raman pump to exceed the ionization limit and hence produce stronger reflection. However, if the plasma frequency continuously increases when the strong peak of the pump pulse arrives, there will be a frequency upshift~\cite{wilks1988frequency, mendoncca2000book, Kenan_2018_upshift,Manuscript_Chirped} and the frequency mismatch may or may not be tolerable for Raman amplification. 
%Moreover, the partially ionized electrons are still susceptible to the periodic background electrostatic potential which may cause strong wave damping as we explained. 

Another method to reduce the plasma density is by ionizing only the surface of the semiconductor or ionizing a thin semiconductor foil and let the plasma expand. However, this method is associated with other issues such as timing of the laser pulses and the wave damping during the plasma expansion. Detailed modeling and numerical simulations are needed to exploit more complicated scenarios.

Our analysis assumes that the onset of ionization is sudden. In practice, ionization time of solid-state semiconductors can range from a single laser cycle when using photoionization to multiple laser cycles when using collisional ionization. If the ionization process is spread into several plasma cycles, the ionized electrons will join the plasma oscillation with different initial phases, which may reduce the coherence and hence the amplitude of the generated plasma wave. However, this restriction is not catastrophic because the plasma oscillation is much slower than the laser electromagnetic oscillation. Even if the ionization is as slow as a few plasma periods, effects of dephasing can be tolerated considering that the plasma frequency only increases gradually so the initial stage accumulates very little phase mismatch.

In principle, for a semiconductor whose lattice constant is near the pump laser wavelength, the lattice structure could be advantageous~\cite{MalkinPRL2007} for improving Raman amplification. Assuming the lattice structure survives during ionization, the plasma concentration is then periodically modulated in space to form a band gap. This band gap, on one hand, can provide an ultrahigh dispersion which effectively stretches the pulse duration and reduces the pulse intensity to delay the self-phase modulation of the pulse; on the other hand, it can forbid the propagation of the frequency-downshifted near-forward Stokes generation, thereby increasing the allowed amplification distance. 
%Note that while the lattice structure might be useful in replacing the plasma channel where the Raman amplification takes place, it should be not used for creating localized plasma waves due to the inhomogeneity induced damping. 

In another speculative application, the creation of localized plasma wave might also offer a new route for inertia confinement fusion (ICF). One can use ionizing semiconductors to create a plasma wave with total energy $\mathcal{E}$ in volume $\mathcal{V}$ and then compress the plasma volume adiabatically using, \eg lasers. According to Refs.~\cite{Schmit_PRL2010, Schmit_PRL2012}, the mechanical energy of compression is partially transferred into the plasma wave leading to increasing the wave energy $\mathcal{E} {\propto} \mathcal{V}^{-1/2}$. The compression also increases the plasma wave vector $k{\propto} \mathcal{V}^{-1}$ until Landau damping suddenly dominates and damps the wave energy into hot electrons. The semiconductor plasma wave is particularly attractive because both the plasma and the embedded wave are well localized, making it easy for compression. It avoids the difficulty of laser energy penetration into overdense plasma targets in current laser heating ICF experiments.

To summarize, the production of plasma waves stands to be useful in a variety of applications. Producing the plasma wave via laser parametric processes is not always possible. We identified how the ionization of doped semiconductors creates an interesting alternative. Although a number of practical issues in applying this technique are addressed here, there remain many further issues to address and further embodiments envisioned before the full potential of this mechanism might be realized.

\begin{acknowledgments}
	The authors thank Prof. A. Kahn (Princeton University) for useful discussions on the details of possible experimental realization. 
	This work was supported by NNSA Grant No. DE-NA0003871, and AFOSR Grant No. FA9550-15-1-0391. The EPOCH code was developed as part of the UK EPSRC grants EP/G054950/1, EP/G056803/1, EP/G055165/1 and EP/ M022463/1.
\end{acknowledgments}  

\bibliography{LangmuirSeed}
% ****** End of file apssamp.tex ******

\end{document}